\documentclass[12pt]{iopart}
\usepackage{graphicx}% Include figure files
\usepackage{dcolumn}% Align table columns on decimal point
\usepackage{bm}% bold math
\usepackage{amssymb}

\renewcommand{\vec}[1]{\ensuremath{\boldsymbol{\mathrm{#1}}}}
\begin{document}

\title[Spin correlations in elastic $e^{+}e^{-}$ scattering in QED]{Spin correlations in elastic $e^{+}e^{-}$ scattering in QED}

\author{N Yongram}

\address{Fundamental Physics \& Cosmology Research Unit\\
The Tah Poe Academia Institute (TPTP), Department of Physics\\
Naresuan University, Phitsanulok, 65000, Thailand}
\ead{nattapongy@nu.ac.th}
\begin{abstract}
Spin correlations are carefully investigated in elastic $e^{+}e^{-}$ scattering in QED, for initially \textit{polarized} as well as \textit{unpolarized} particles,
with emphasis placed on energy or speed of the underlying particles involved in the process. An explicit expression is derived for the corresponding transition probabilities in closed form to the leading order. These expressions are unlike the ones
obtained from simply combining spins of the relevant particles which are of kinematic nature.
It is remarkable that these explicit results obtained from quantum field theory show a clear violation of Bell's inequality at \textit{all} energies in support of
quantum theory in the relativistic regime. We hope that our explicit expression obtained will lead to experiments in the manner described in the bulk of this paper by monitoring speed.

\end{abstract}

\pacs{12.20.Ds,  12.20.Fv, 42.50.Wm, 03.65.Bz.}
\maketitle

\section{Introduction}
In recent years, we have been interested in studying joint polarization correlations of fundamental processes in QED and in the electro-weak theory\cite{NY_EBM;2003,EBM_NY;2004,EBM_NY;2005,NY_EBM_SS;2006} for initially \textit{polarized} and \textit{unpolarized} particles. Our main conclusion, based on explicit computations in quantum field theory, is that the mere
fact that particles emerging from a process have non-zero speeds to reach detectors implies, in general, that their spin polarizations correlations probabilities
\textit{depend on speed} \cite{NY_EBM;2003,EBM_NY;2004,EBM_NY;2005,NY_EBM_SS;2006} and \textit{may also depend on the underlying couplings} \cite{NY_EBM_SS;2006}. The explicit expressions of polarization correlations, follow from these \textit{dynamical} computations and are non-speculative
involving no arbitrary input assumptions, and are seen
to depend on speed, and may depend on the couplings as well. These are unlike na\"{\i}ve considerations of simply combining
the spins of the particles in question which are of kinematical
nature, in general, not applicable to relativistic particles, and formal arguments based on combining spins only, as is often done, completely fail.
In the limit of low energies, our earlier expressions \cite{NY_EBM;2003,EBM_NY;2004,EBM_NY;2005,NY_EBM_SS;2006} for the polarization correlations were shown to be reduced to the na\"{\i}ve ones just mentioned by simply combining spins. In our earlier investigation \cite{NY_EBM_SS;2006} in the electro-weak theory in a process for the creation of a $\mu^{+}\mu^{-}$ pair, from $e^{+}e^{-}$ scattering, for example it was noted that
due to the threshold needed to create such a pair the zero energy limit may not be taken and the study of polarization correlations by simply combining spins,
without recourse to quantum field theory, has no meaning. Our interest in this paper is the derivation of the \textit{explicit} polarization correlation probabilities in elastic $e^{+}e^{-}$ scattering in QED for initially \textit{polarized} as well as \textit{unpolarized} particles, with emphasis put on the \textit{energy} available in the process so that a detailed study can be carried out in the relativistic regime as well. The reasons for our present investigation are two fold. First several experiments on $e^{+}e^{-}\to{}e^{+}e^{-}$ have been carried out over the years \cite{Howe;1953,Ashkin;1954,Augustin;1975,Learned;1975,O'Neill;1976}, and it is expected that our explicit new expression for the polarization correlations obtained, depending on speeds, may lead to new experiments on polarization correlations which monitor the speed of the underlying particles. Second, such a study may be relevant to experiments in the light of Bell's theorem (monitoring speed) as discussed below.

The relevant quantity of interest here in testing Bell's
inequality \cite{Clauser_Horne;1974,Clauser_Shimoney;1978} is, in a standard
notation,
\begin{eqnarray}
  S &= \frac{p_{12}(a_{1},a_{2})}{p_{12}(\infty,\infty)}
  -\frac{p_{12}(a_{1},a'_{2})}{p_{12}(\infty,\infty)}
  +\frac{p_{12}(a'_{1},a_{2})}{p_{12}(\infty,\infty)}
  +\frac{p_{12}(a'_{1},a'_{2})}{p_{12}(\infty,\infty)}
  \nonumber \\[0.5\baselineskip]
  &\quad
  -\frac{p_{12}(a'_{1},\infty)}{p_{12}(\infty,\infty)}
  -\frac{p_{12}(\infty,a_{2})}{p_{12}(\infty,\infty)}
  \label{Eq01}
\end{eqnarray}
as is \emph{computed from} QED.   Here $a_{1}$,
$a_{2}$ \ $(a'_{1},a'_{2})$ specify directions along which the
polarizations of two particles are measured, with
$p_{12}(a_{1},a_{2})/p_{12}(\infty,\infty)$ denoting the joint
probability, and $p_{12}(a_{1},\infty)/p_{12}(\infty,\infty)$,\
$p_{12}(\infty,a_{2})/p_{12}(\infty,\infty)$ denoting the
probabilities when the polarization of only one of the particles
is measured.   [$p_{12}(\infty,\infty)$ is normalization factor.]
The corresponding probabilities as computed from QED
will be denoted by $P[\chi_{1},\chi_{2}]$, $P[\chi_{1},-]$,
$P[-,\chi_{2}]$ with $\chi_{1}$, $\chi_{2}$ denoting angles
specifying directions along which spin measurements are carried
out with respect to certain axes spelled out in the bulk of the
paper. To show that the QED process is in violation with
Bell's inequality of LHV, it is sufficient to find one set of
angles $\chi_{1}$, $\chi_{2}$, $\chi'_{1}$, $\chi'_{2}$, such that
$S$, as computed in QED, leads to a value of $S$
outside the interval $[-1,0]$. In this work, it is implicitly
assumed that the polarization parameters in the particle states
are directly observable and may be used for Bell-type measurements
as discussed. We show a clear violating of Bell's inequality for \textit{all} speeds in support of
quantum theory in the relativistic regime, i.e., of quantum field theory.

\section{Spin correlations; initially polarized particles}

\begin{figure}
\centering
\includegraphics[width=0.5\textwidth]{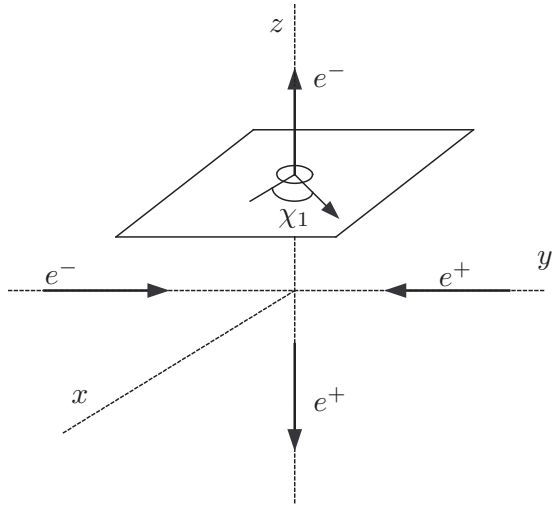}% Here is how to import EPS art
\caption{ The figure depicts $e^{+}e^{-}$ scattering, with the electron and positron
initially moving the $y$-axis, while the emerging electron and positron moving along the $z$-axis.
The angle $\chi_{1}$, measured relative to the $x$-axis, denotes the orientation of spin of the emerging electron
may made.}\label{fig_1}
\end{figure}
We consider the process  $e^{+}e^{-}\to{}e^{+}e^{-}$ in the center of mass frame (see Fig.~\ref{fig_1}), with initially polarized electron-positron with one spin up, along $z$-axis, and one spin down. With $\vec{p}_{1}=\gamma{}m\beta(0,1,0)=-\vec{p}_{2}$  denoting the momenta of initial electron and positron, respectively, $\gamma=1/\sqrt{1-\beta^{2}}$, we consider momenta of emerging electron and positron with
\begin{equation}
\vec{k}_{1}=\gamma{}m\beta(0,0,1)=-\vec{k}_{2} \label{Eq02}
\end{equation}
For four-spinors of the initial electron and positron, respectively, we have
\begin{eqnarray}
u(p_{1})&\sim\left(
  \begin{array}{c}
    \uparrow \\
    \mathrm{i}\rho\downarrow
  \end{array}
\right)
\;\;\mathrm{and}\;\;
\overline{v}(p_{2})\sim\left(
  \begin{array}{cc}
    \mathrm{i}\rho\uparrow^{\dagger} & -\downarrow^{\dagger}
  \end{array}
\right)\label{Eq03}\\[0.5\baselineskip]
\rho&=\frac{\gamma\beta}{\gamma+1}=\frac{\beta}{1+\sqrt{1-\beta^{2}}} \label{Eq04}
\end{eqnarray}
where $\uparrow\equiv\left(
  \begin{array}{c}
    1 \\
    0
  \end{array}
\right)$ is a spin up,  $\downarrow\equiv\left(
  \begin{array}{c}
    0 \\ 1
  \end{array}
\right)$ is a spin down, $\uparrow^{\dagger}\equiv\left(
  \begin{array}{cc}
    1 & 0
  \end{array}
\right)$ is a transpose matrix of spin up and  $\downarrow^{\dagger}\equiv\left(
  \begin{array}{cc}
   0 & 1
  \end{array}
\right)$ is a  transpose matrix of spin down, and for four-spinor of emerging electron and positron, respectively, we have
\begin{eqnarray}
\overline{u}(k_{1})&\sim\left(
  \begin{array}{cc}
    \zeta^{\dagger}_{1} & \rho\sigma_{3}\zeta^{\dagger}_{1}
  \end{array}
\right)\;\;\mathrm{and}\;\;
v(k_{2})\sim\left(
  \begin{array}{c}
    \rho\sigma_{3}\zeta_{2} \\ \zeta_{2}
  \end{array}
\right)\label{Eq05}\\[0.5\baselineskip]
\rho&=\frac{\gamma\beta}{\gamma+1}=\frac{\beta}{1+\sqrt{1-\beta^{2}}} \label{Eq06}
\end{eqnarray}
where the two-spinors  $\zeta_{1}$, $\zeta_{2}$   will be specified later.

The expression for the amplitude of the process is well known \cite{Itzykson;1980,Griffiths}
\begin{eqnarray}
\mathcal{M}&\propto\overline{v}(p_{2})\gamma^{\mu}u(p_{1})\overline{u}(k_{1})
\gamma_{\mu}v(k_{2})\frac{1}{(p_{1}+p_{2})^{2}}\nonumber\\[0.5\baselineskip]
&-\overline{u}(k_{1})\gamma^{\mu}u(p_{1})\overline{v}(p_{2})
\gamma_{\mu}v(k_{2})\frac{1}{(p_{1}-k_{1})^{2}} \label{Eq07}
\end{eqnarray}
Given that such as  the amplitude of the process above, we compute the conditional joint probability
of spins measurements of $e^{+}$, $e^{-}$ along directions specified by the angles $\chi_{1}$, $\chi_{2}$
as shown in Fig.~\ref{fig_1}. We then have two-spinors as
\begin{eqnarray}
\zeta_{1}=\case{1}{\sqrt{2}}\left(
  \begin{array}{c}
    e^{-\mathrm{i}\chi_{1}/2} \\ e^{\mathrm{i}\chi_{1}/2}
  \end{array}
\right)
\;\mathrm{and}\; \zeta_{2}=\case{1}{\sqrt{2}}\left(
  \begin{array}{c}
    e^{-\mathrm{i}\chi_{2}/2} \\ e^{\mathrm{i}\chi_{2}/2}
  \end{array}
\right)\label{Eq08}
\end{eqnarray}
Here we have considered only the single state cf. \cite{EBM_NY;2004,NY_EBM_SS;2006}. By using the four-spinors of initial and emerging
 $e^{+}$, $e^{-}$, and  two-spinors in term of the angles $\chi_{1}$, $\chi_{2}$ to calculate the invariant amplitude of the process in
 Fig.~\ref{fig_1} leads to
\begin{eqnarray}
\mathcal{M}&\propto\left[A(\beta)\cos\left(\frac{\chi_{1}+\chi_{2}}{2}\right)
+B(\beta)\sin\left(\frac{\chi_{1}-\chi_{2}}{2}\right)\right]\nonumber\\[0.5\baselineskip]
&+\mathrm{i}\left[C(\beta)\sin\left(\frac{\chi_{1}+\chi_{2}}{2}\right)
+D(\beta)\cos\left(\frac{\chi_{1}-\chi_{2}}{2}\right)\right] \label{Eq09}
\end{eqnarray}
where
\begin{eqnarray*}
A(\beta)&=1-\rho^{2}(1-\rho)+2\beta^{2}(1-\rho^{2})^{2}\\[0.5\baselineskip]
B(\beta)&=\rho(1+\rho)+8\beta^{2}\rho^{2} \\[0.5\baselineskip]
C(\beta)&=1+\rho^{2}(1-\rho)+2\beta(1-\rho^{4}) \\[0.5\baselineskip]
D(\beta)&= \rho(1+\rho)
\end{eqnarray*}
Using the notation $F[\chi_{1},\chi_{2}]$ for the absolute value square of the right-hand side of (\ref{Eq09}),
the conditional joint probability distribution of spin measurements along the directions specified by angle
$\chi_{1}$, $\chi_{2}$ is given by
 \begin{equation}
P[\chi_{1},\chi_{2}]=\frac{F[\chi_{1},\chi_{2}]}{N(\beta)} \label{Eq10}
\end{equation}
The normalization factor $N(\beta)$ is obtained by summing over all the polarizations of the emerging particles. This is
equivalent to summing of $F[\chi_{1},\chi_{2}]$ over the pairs of angles
 \begin{eqnarray}
(\chi_{1},\chi_{2}),\quad(\chi_{1}+\pi,\chi_{2}),\quad(\chi_{1},\chi_{2}+\pi),\quad(\chi_{1}+\pi,\chi_{2}+\pi) \label{Eq11}
\end{eqnarray}
leads to
\begin{eqnarray}
N(\beta)&=F[\chi_{1},\chi_{2}]+F[\chi_{1}+\pi,\chi_{2}]\nonumber\\[0.5\baselineskip]
&+F[\chi_{1},\chi_{2}+\pi]+P[\chi_{1}+\pi,\chi_{2}+\pi]\nonumber\\[0.5\baselineskip]
&=2[A^{2}(\beta)+B^{2}(\beta)+C^{2}(\beta)+D^{2}(\beta)] \label{Eq12}
\end{eqnarray}
giving
\begin{eqnarray}
P[\chi_{1},\chi_{2}]&=\frac{\left[A(\beta)\cos\left(\frac{\chi_{1}+\chi_{2}}{2}\right)
+B(\beta)\sin\left(\frac{\chi_{1}-\chi_{2}}{2}\right)\right]^{2}}{2[A^{2}(\beta)+B^{2}(\beta)+C^{2}(\beta)+D^{2}(\beta)]}\nonumber\\[0.5\baselineskip]
&+\frac{\left[C(\beta)\sin\left(\frac{\chi_{1}+\chi_{2}}{2}\right)
+D(\beta)\cos\left(\frac{\chi_{1}-\chi_{2}}{2}\right)\right]^{2}}{2[A^{2}(\beta)+B^{2}(\beta)+C^{2}(\beta)+D^{2}(\beta)]} \label{Eq13}
\end{eqnarray}
If only one of the spins is measured, say, corresponding to $\chi_{1}$, we then have the probability as
\begin{eqnarray}
P[\chi_{1},-]&=P[\chi_{1},\chi_{2}]+P[\chi_{1},\chi_{2}+\pi]\nonumber\\[0.5\baselineskip]
               &=\frac{1}{2}+\frac{2[A(\beta)B(\beta)+C(\beta)D(\beta)]\sin\chi_{1}}{2[A^{2}(\beta)+B^{2}(\beta)+C^{2}(\beta)+D^{2}(\beta)]} \label{Eq14}
\end{eqnarray}
Similarly, the only one of the spins is measured corresponding to $\chi_{2}$, we then have the probability as
\begin{eqnarray}
P[-,\chi_{2}]&=P[\chi_{1},\chi_{2}]+P[\chi_{1}+\pi,\chi_{2}]\nonumber\\[0.5\baselineskip]
               &=\frac{1}{2}+\frac{2[C(\beta)D(\beta)-A(\beta)B(\beta)]\sin\chi_{2}}{2[A^{2}(\beta)+B^{2}(\beta)+C^{2}(\beta)+D^{2}(\beta)]}
               \label{Eq15}
\end{eqnarray}
For all $0\leqslant\beta\leqslant1$, angles $\chi_{1}$, $\chi_{2}$, $\chi'_{1}$, $\chi'_{2}$ are readily found leading to
a violation of Bell's inequality of LHV theories. For example, for $\beta=0.9$, $\chi_{1}=0^{\circ}$, $\chi_{2}=45^{\circ}$,
 $\chi'_{1}=69^{\circ}$, $\chi'_{2}=200^{\circ}$, $S=-1.311$ violating the inequality from below.

\section{Spin correlations; initially unpolarized particles}
For the process $e^{+}e^{-}\to{}e^{+}e^{-}$, in the center of mass (c.m.), with initially unpolarized spins, with momenta
$\vec{p}_{1}=\gamma{}m\beta(0,1,0)=-\vec{p}_{2}$,
\begin{figure}
\centering
\includegraphics[width=0.6\textwidth]{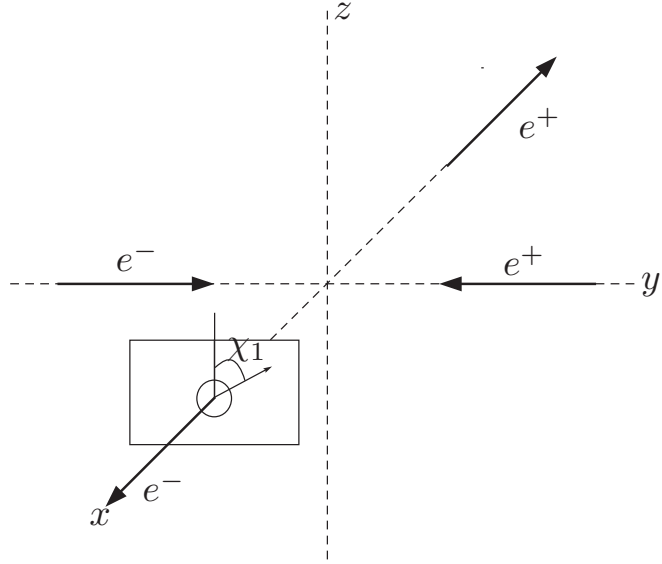}% Here is how to import EPS art
\caption{\label{fig_2} The figure depicts $e^{+}e^{-}$ scattering, with $e^{+}$, $e^{-}$ moving along the $y$-axis,
and the emerging electron and positron moving along the $x$-axis.
The angle $\chi_{1}$, measured relative to the $z$-axis, denotes the orientation of spin of the emerging electron
may made.}
\end{figure}
we take for the final electron and positron
\begin{eqnarray}
\vec{k}_{1}=\gamma{}m\beta(1,0,0)=-\vec{k}_{2}\label{Eq16}
\end{eqnarray}
and for the four-spinors
\begin{eqnarray}
u(k_{1})&=\left(\frac{k^{0}+m}{2m}\right)^{\!\!1/2}\!\!\left(
  \begin{array}{c}
  \xi_{1} \\ {}\frac{\vec{k}_{1}\cdot\vec{\sigma}}{k^{0}+m}\xi_{1}
  \end{array}
\right),\;
\xi_{1}=\left(
  \begin{array}{c}
    -\mathrm{i}\cos\chi_{1}/2\\ \sin\chi_{1}/2
  \end{array}
\right)\label{Eq17}\\[0.5\baselineskip]
v(k_{2})&=\left(\frac{k^{0}+m}{2m}\right)^{\!\!1/2}\!\!\left(
  \begin{array}{c}
    \frac{\vec{k}_{2}\cdot\vec{\sigma}}{k^{0}+m}\xi_{2} \\ {}\xi_{2}
  \end{array}
\right),\;
\xi_{2}=\left(
  \begin{array}{c}
     -\mathrm{i}\cos\chi_{2}/2\\ \sin\chi_{2}/2
  \end{array}
\right)\label{Eq18}
\end{eqnarray}
The absolute value square of the right-hand side of (\ref{Eq09}) with initially unpolarized electrons and positron leads
to
\begin{eqnarray}
\mathrm{Prob}&\propto\frac{\Tr\left[\gamma^{\sigma}(\gamma{}p_{2}+m)\gamma^{\mu}(-\gamma{}p_{1}+m)\right]
\overline{u}(k_{1})\gamma_{\mu}v(k_{2})\overline{v}(k_{2})\gamma_{\sigma}u(k_{1})}{(p_{1}+p_{2})^{4}}\nonumber\\[0.5\baselineskip]
&-\frac{\Tr\left[(\gamma{}p_{2}+m)\gamma^{\mu}(-\gamma{}p_{1}+m)\gamma^{\sigma}\right]
\overline{v}(k_{2})\gamma_{\sigma}u(k_{1})\overline{u}(k_{1})\gamma_{\mu}v(k_{2})}{(p_{1}+p_{2})^{2}(p_{1}-k_{1})^{2}}\nonumber\\[0.5\baselineskip]
&-\frac{\Tr\left[\gamma^{\mu}(-\gamma{}p_{1}+m)\gamma^{\sigma}(\gamma{}p_{2}+m)\right]
\overline{u}(k_{1})\gamma_{\mu}v(k_{2})\overline{v}(k_{2})\gamma_{\sigma}u(k_{1})}{(p_{1}+p_{2})^{2}(p_{1}-k_{1})^{2}}\nonumber\\[0.5\baselineskip]
&+\frac{\overline{u}(k_{1})\gamma^{\mu}(-\gamma{}p_{1}+m)\gamma^{\sigma}u(k_{1})
\overline{v}(k_{2})\gamma_{\sigma}(\gamma{}p_{2}+m)\gamma_{\mu}v(k_{2})}{(p_{1}-k_{1})^{4}}\label{Eq19}
\end{eqnarray}
which after simplification and of collecting terms reduces to
\begin{eqnarray}
 \mathrm{Prob}&\propto[2\beta^{4}(1+2\beta^{2})-3(1+\beta^{2})]\sin^2\left(\case{\chi_1-\chi_2}{2}\right)
 \nonumber\\[0.5\baselineskip]
 &\quad+(1+\beta^{2}+2\beta^{4})\cos^2\left(\case{\chi_1+\chi_2}{2}\right)+5(1-\beta^{2})\nonumber\\[0.5\baselineskip]
 &\equiv F[\chi_1,\chi_2] \label{Eq20}
\end{eqnarray}
where we have used the expressions for the spinors in
(\ref{Eq17}), (\ref{Eq18}).

Given that the process has occurred, the conditional probability
that the spins of the emerging electron and positron make angles $\chi_1$,
$\chi_2$, respectively, with the $z$-axis, is directly obtained from
(\ref{Eq20}) to be
\begin{eqnarray}
 P[\chi_1,\chi_2]=\frac{F[\chi_1,\chi_2]}{C} \label{Eq21}
\end{eqnarray}
The normalization constant $C$ is obtained by summing over the
polarizations of the emerging electrons. This is equivalent to
summing of $F[\chi_1,\chi_2]$ over the pairs of angles in (\ref{Eq11})
for any arbitrarily chosen fixed $\chi_1$, $\chi_2$, corresponding
to the orthonormal spinors
\begin{eqnarray}
\left(
  \begin{array}{c}
  -\mathrm{i}\cos\chi_j/2\\
  \sin\chi_j/2
  \end{array}
\right),\quad\left(
  \begin{array}{c}
  -\mathrm{i}\cos(\chi_j+\pi)/2\\
  \sin(\chi_j+\pi)/2
  \end{array}
\right)=\left(
  \begin{array}{c}
    \mathrm{i}\sin\chi_j/2\\
  \cos\chi_j/2
  \end{array}
\right)\label{Eq22}
\end{eqnarray}
providing a complete set, for each $j=1,2$, in reference to
(\ref{Eq17}), (\ref{Eq18}).This is,
\begin{eqnarray}
 C&=F[\chi_1,\chi_2]+F[\chi_1+\pi,\chi_2]\nonumber\\[0.5\baselineskip]
 &\quad+F[\chi_1,\chi_2+\pi]+F[\chi_1+\pi,\chi_2+\pi]
 \nonumber\\[0.5\baselineskip]
 &=8[2-3\beta^{2}+\beta^{4}+\beta^{6}] \label{Eq23}
\end{eqnarray}
which as expected is independent of $\chi_1$, $\chi_2$, giving
\begin{eqnarray}
 P[\chi_1,\chi_2]&=\frac{[2\beta^{4}(1+2\beta^{2})-3(1+\beta^{2})]\sin^2\left(\frac{\chi_1-\chi_2}{2}\right)}{8[2-3\beta^{2}+\beta^{4}+\beta^{6}]}
 \nonumber\\[0.5\baselineskip]
 &\quad+\frac{[1+\beta^{2}+2\beta^{4}]\cos^2\left(\frac{\chi_1+\chi_2}{2}\right)+5(1-\beta^{2})}{8[2-3\beta^{2}+\beta^{4}+\beta^{6}]}\label{Eq24}
\end{eqnarray}
By summing over
\begin{eqnarray}
\chi_2,\quad \chi_2+\pi \label{Eq25}
\end{eqnarray}
for any arbitrarily fixed $\chi_2$, we obtain
\begin{eqnarray}
P[\chi_1,-]=\frac{1}{2} \label{Eq26}
\end{eqnarray}
and similarly,
\begin{eqnarray}
P[-,\chi_2]=\frac{1}{2} \label{Eq27}
\end{eqnarray}
for the probabilities when only one of the photons polarizations
is measured.

A clear violation of Bell's inequality of LHV theories was
obtained for all $0\leqslant\beta\leqslant1$. For example,
for $\beta=0.8$, with $\chi_1=0^{\circ}$, $\chi_2=45^{\circ}$,
$\chi'_1=210^{\circ}$, $\chi'_2=15^{\circ}$ give $S=-1.167$
violating the inequality from below.
violates LHV theories.\\

\section{Conclusion}
A critical study of polarization correlations probabilities in elastic $e^{+}e^{-}$ scattering was carried out, for initially \textit{polarized} as well as \textit{unpolarized} particles, emphasizing their dependence on speed and an explicit expression for them obtained in QED. The necessity of such a study within the realm of quantum field theory cannot be overemphasized as estimates of such correlations obtained by simply combining spins, as is often done, have no meaning as they do not involve dynamical considerations. The relevant dynamics is, of course, dictated directly from quantum field theory. The explicit expression for the polarization correlation obtained is interesting in its own right but may also lead to experiments which investigate such correlations by monitoring speed not only for initially polarized particles but also for unpolarized ones. Our results may be also relevant in the realm of Bell's inequality with emphasis put on relativistic aspects of quantum theory, that is of quantum field theory. Our expressions have shown clear violations of Bell's inequality of LHV theories in support of quantum theory in the relativistic regime. Several experiments have been already performed in recent years (cf.
\cite{Aspect_Dalibard_Roger;1982,Irby;2003,Osuch;1996,Kaday_Ulman_Wu;1975,Fry;1995}) on particles' polarizations correlations. And, it is expected that the novel properties recorded here by explicit calculations following directly from field theory, which is based on the principle of relativity and quantum theory, will lead to new experiments on polarization correlations monitoring speed in the light of Bell's Theorem. We hope that theses computations, in general setting of quantum field theory, will be also useful in such areas of physics as quantum teleportation and quantum information in general.

\section*{Acknowledgments}
I would like to thank Prof. Dr. E. B. Manoukian for discussions, guidance and for carefully reading the manuscript. I also would like to thank Suppiya Siranan and Dr. Burin Gumjudpai for discussions and comments.
Finally, I would like to acknowledge with thanks for the award of a research grant by the TRF-New Researcher Grant of the Thailand Research
Fund (MRG5080288) and my gratitude to the Faculty of Science, Naresuan University.

\end{document}